%% file: main.tex
\documentclass[twocolumn,tighten]{aastex702}

\usepackage{CJKutf8}

\input{values.tex}

\begin{document}
\begin{CJK*}{UTF8}{gbsn}

\title{The Impact of an AGN on PAH Spectral Features in Galaxy Evolution Simulations: Continuum Dilution and Excitation of PAHs}

\author[0009-0001-6065-0414]{Grant P. Donnelly}
\affiliation{Ritter Astrophysical Research Center, University of Toledo, Toledo, OH 43606, USA}
\email[show]{Grant.Donnelly@rockets.utoledo.edu}
\correspondingauthor{Grant P. Donnelly}

\author[0000-0002-7064-4309]{Desika Narayanan}
\affiliation{Department of Astronomy, University of Florida, 211 Bryant Space Sciences Center, Gainesville, FL 32611 USA}
\affiliation{Cosmic Dawn Center at the Niels Bohr Institute, University of Copenhagen and DTU-Space, Technical University of Denmark}
\email{desika.narayanan@ufl.edu}

\author[0000-0002-5653-0786]{Paul Torrey}
\email{ygn5rz@virginia.edu}
\affiliation{Department of Astronomy, University of Virginia, 530 McCormick Road, Charlottesville, VA 22904}
\affiliation{Virginia Institute for Theoretical Astronomy, University of Virginia, Charlottesville, VA 22904, USA}
\affiliation{The NSF-Simons AI Institute for Cosmic Origins, USA}

\author[0000-0003-1545-5078]{J.-D. T. Smith}
\affiliation{Ritter Astrophysical Research Center, University of Toledo, Toledo, OH 43606, USA}
\email{JD.Smith@utoledo.edu}

\author[0000-0001-6325-9317]{Helena M. Richie}
\affiliation{Center for Computational Astrophysics, Flatiron Institute, 162 5th Avenue, New York, NY 10010, USA}
\email{helenarichie@gmail.com}

\author[0000-0002-1253-2763,gname=Hui,sname=Li]{Hui Li (李辉)}
\affiliation{Department of Astronomy, Tsinghua University, Beijing, 100084, People's Republic of China}
\email{hliastro@tsinghua.edu.cn}

\author[0000-0003-3816-7028]{Federico Marinacci}
\affiliation{Department of Physics and Astronomy ``Augusto Righi'', University of Bologna, Via P. Gobetti 93/2, I-40129 Bologna, Italy}
\affiliation{INAF, Astrophysics and Space Science Observatory Bologna, Via P. Gobetti 93/3, I-40129 Bologna, Italy}
\email{federico.marinacci2@unibo.it}

\begin{abstract}
	
Galaxies hosting an active galactic nucleus (AGN) show diminished power and non-standard spectral feature ratios from polycyclic aromatic hydrocarbons (PAHs) compared with normal star-forming galaxies. Proposed mechanisms for this difference include a combination of AGN-driven PAH grain processing, differential heating from the surrounding radiation field, and continuum dilution. These potentially competing factors are difficult to disentangle observationally. In this paper, we investigate the impact of the AGN radiation field on the excitation properties and potential dilution of the mid-IR emission features via constrained numerical experiments of idealized galaxies in simulations that include a self-consistent model for the evolution of dust. We find that, for the integrated spectrum of our simulated Milky Way analog and independent of grain-processing physics, the apparent PAH emission is progressively diminished as AGN luminosity is increased. For spectra within the central half kiloparsec, PAH features are entirely obfuscated when an AGN of any sampled luminosity is present. We additionally find that the UV radiation from an AGN is capable of impacting PAH band ratios, though note that we do not yet include a model for potential photodestruction of PAHs. Our model suggests that the inferred lack of PAHs via featureless mid-IR spectra from galaxies hosting AGN may simply owe to the direct continuum from the AGN overpowering mid-IR features from PAHs in the ISM of the host galaxy. Future work will investigate the relative role of AGN continuum dilution and PAH heating vs photodestruction/ionization of PAHs.
	
\end{abstract}


\section{Introduction}

The mid-infrared (MIR) spectral emission features arising from small interstellar dust grains known as polycyclic aromatic hydrocarbons (PAHs) are altered, diminished, or even absent in galaxies where an active galactic nucleus (AGN) is present \citep[e.g.,][]{aitken_roche_1985, moorwood_1986, smith_07, odowd_09, diamond-stanic_10}, compared to the PAH-dominated global spectra of normal star-forming galaxies where variations are relatively muted. Less affected by extinction than optical/UV diagnostics, PAH emission is an increasingly common indicator for the star-formation rate (SFR) \citep[e.g.,][]{peeters_2004, shipley_2016, lai_2020} and is a promising probe for conditions in the interstellar medium (ISM) such as the molecular gas content \citep{chown_2025} and the gas-phase metallicity \citep{whitcomb_2024}. However, the changes to PAH spectra when an AGN is present make these uses difficult to interpret for those galaxies. For example, reduced PAH emission could lead to underestimates of the SFR, and inferred $\mathrm{H_2}$ quantities could be affected as AGN tend to have larger $\mathrm{H_2 / PAH}$ ratios \citep[e.g.,][]{lambrides_2019}.

It is common to ascribe these different PAH spectra to an influence of the AGN on PAH grains themselves. For instance, the hard radiation emitted by an AGN could destroy PAH grains, leading to an overall reduction in PAH emission \citep{aitken_roche_1985, voit_1992}. For AGN-hosting galaxies in which PAH emission is observed, short-to-long (STL) wavelength PAH ratios are suppressed \citep[e.g.,][]{smith_07, odowd_09, sales_10, garciabernete_22, lai_2022}, possibly indicating that an AGN preferentially destroys the smaller (and perhaps more brittle) PAH grains, which tend to emit more power at shorter wavelengths \citep{draine_2021}. If this is true, the inference is that AGN may play a significant role in the evolution of the dust grain size distribution (GSD) and total abundance of dust within a galaxy. However, the details surrounding this mechanism are unclear, complicated in part by evidence that the very smallest grains can still emit brightly in AGN hosts \citep{lai_2023}. Simultaneously, the shape of the interstellar radiation field (ISRF) spectrum that excites PAHs also impacts PAH emission by partly determining their temperature. As an AGN impacts the ISRF of a galaxy through its influence on the star formation history (SFH), an AGN may indirectly influence PAH emission even when there is no physical change to grain properties.

Beyond the potential physical impact of AGN on PAHs, the direct emission from the dusty torus structure can contribute significantly to the mid-IR SED \citep{pier_1992,Nenkova_I,Nenkova_II}, and potentially dilute PAH features. Indeed, this dilution is often exploited as an indicator for an AGN in obscured environments  \citep[e.g.,][]{genzel_1998, laurent_2000, tran_2001, armus_2007, valiante_2007, rich_2023}. Further, since individual PAH features have different widths and relative strengths, the sloped AGN continuum could differentially affect PAH band ratios as it dilutes the PAH features. The apparent lack of PAH emission reported by early, necessarily large-aperture, MIR studies of local AGN-hosts was likely influenced by this dilution effect. With JWST or Spitzer, continuum dilution effects are minimal in the extended regions of nearby resolved galaxies, but this effect will still hinder the study of PAHs in the nucleus where any direct impact by the AGN on PAHs would be the most significant. On the other hand, the observed SEDs of unresolved, distant AGN-hosts and quasars include a blend of AGN and host galaxy components by definition. This can be a significant source of uncertainty for current studies of PAHs at high redshift, and will be a source of uncertainty for future far-infrared facilities, such as PRIMA, as they use PAHs to track the accumulation of metals throughout cosmic time.

The efficacy of PAH emission as a SFR tracer owes to the fact that PAHs are more selective of UV/optical photons \citep{allamandola_1989, DL_2007, draine_2021}. As AGN emit substantial power at these wavelengths due to thermal emission from the accretion disk \citep[e.g.,][]{malkan_1982, marchese_2012}, an AGN may be able to supply a substantial fraction of photons for PAH excitation for grains surviving in these galaxies. Indeed, it has been shown using high spatial resolution ground-based MIR imaging that PAH surface brightness within the centers of galaxies actually increases as the AGN is approached \citep{jensen_2017}. Thus, in star-forming systems that host an AGN, it is critical to understand whether and how an AGN can power PAH emission in order to properly interpret how PAH luminosity traces the SFR, molecular gas content, or gas-phase metallicity. Additionally, PAH band ratios are commonly used to trace conditions in the ISM because these ratios are sensitive to factors like the PAH GSD, PAH charge, and PAH temperature \citep{draine_2021}. As an AGN influences the shape of the ISRF that excites PAHs by boosting the fraction of high-energy photons, the AGN also has the potential to influence PAH band ratios through heating effects only \citep{donnelly_2024}; PAHs receiving more high energy photons will tend to emit with higher STL ratios. This effect would compete with the STL ratio-suppressing impact of AGN-driven small grain destruction.

Observationally, it is difficult to separate these competing effects: continuum dilution, possible AGN-driven PAH grain destruction, and the blend of radiation fields from sources that excite PAH emission. Several studies have modeled the impact of PAH-AGN continuum dilution through mixing models \citep[e.g.,][]{lutz_1998, xie_ho_2022}, but so far each of these have necessarily treated the AGN and host galaxy (including PAH) SED components as separable and independent. However, such methods cannot account for the possibility of PAH excitation from AGN heating, and they are not applicable for off-nuclear apertures, where ISM heating may still be influenced by the AGN. Studying these effects in simulated galaxies provides a path forward. The modern radiative transfer code \texttt{POWDERDAY} \citep{POWDERDAY}, combined with the theoretical PAH emission models of \cite{draine_2021} and \citet{richie_2026}, has enabled the study of simulated PAH spectra that are sensitive to changes in the radiation field hardness and intensity, as well as the grain size and ionization state distributions. 

The purpose of this work is to conduct a numerical experiment using \texttt{POWDERDAY} to include an AGN as a radiation component for PAHs in a simulated galaxy, and to examine how continuum dilution and grain heating affects PAH emission in the resulting galaxy SED in the context of self-consistent radiative transfer. This is a necessary step for future studies to properly interpret the SEDs of simulated galaxies that more fully model AGN-driven PAH processing and the geometry-dependent anisotropy of AGN radiation. This work is organized as follows: in Section \ref{sec:simulations_description}, we describe our galaxy evolution simulation, the radiative transfer numerical experiment with \texttt{POWDERDAY}, and how we include an AGN in it. In Section \ref{sec:results}, we describe the results of the numerical experiment, and in Section \ref{sec:discussion} we discuss the implications of these results on PAH visibility and excitation in AGN-host galaxies. Finally, we summarize this work in Section \ref{sec:summary}.

\section{Simulated AGN-Host Galaxy Spectra} \label{sec:simulations_description}

Our goal is to produce simulated MIR spectra of an AGN-host, so we must first produce a simulated galaxy. In this section, we describe how we generated such a galaxy and modeled the radiative transfer in it.

\subsection{Galaxy Formation and Evolution} \label{sec:galaxy_evo}

\begin{figure*}
	\centering
	\includegraphics[width=0.99\textwidth]{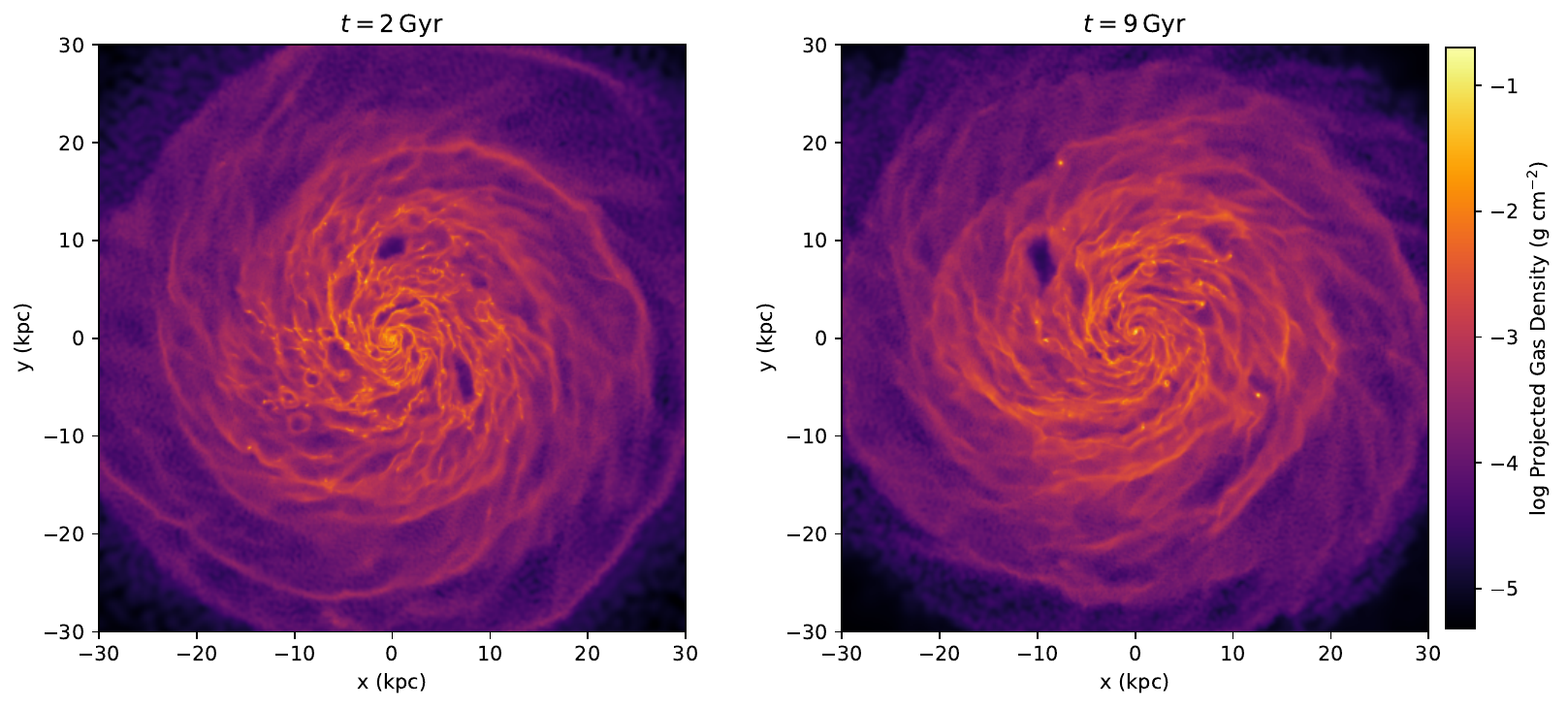}
	\caption{
		Gas density projections along the z-axis for the \texttt{AREPO} simulated Milky Way analog galaxy at $t=2\,\mathrm{Gyr}$ (left) and $t=9\,\mathrm{Gyr}$ (right). In each panel, the AGN is inserted at the center of mass of the gas distribution shown here at (x=0, y=0).
}
	\label{fig:gas_density_plot}
\end{figure*}

In general, our simulation framework is the same as described in \cite[][hereafter N23]{narayanan_2023}.
Specifically, our simulations employ \texttt{AREPO} \citep{AREPO1, AREPO2} with the \texttt{SMUGGLE} module \citep{SMUGGLE} for the galaxy formation and evolution processes (i.e., star formation, cooling, and stellar feedback) coupled with an on-the-fly dust evolution model~\citep{narayanan_2023} that builds upon~\citet{McKinnon2018} and~\citet{Li2019, Li2021} to track dust production, destruction, and the evolution of the dust GSD.
We refer to N23 for a more complete description, and summarize the salient details here. There are two species of dust grains considered, carbonaceous and silicates, and these are spawned stochastically from evolved stars and from supernovae near star-forming regions. In the N23 framework, there is no formal distinction between PAHs and large aromatic carbonaceous grains; all aromatic carbonaceous grains may emit PAH spectral features. However, the larger grains do not contribute significantly to the PAH spectral features as they can not reach the high temperatures required to do so in most environments.
 
PAHs face a variety of hazards in the ISM, and so grain destruction mechanisms are an important part of the PAH lifecycle. In the N23 framework, thermal sputtering and shocks near star-forming regions may destroy PAHs. Grain-grain shattering alters the dust GSD, but conserves the mass within dust. Photodestruction of PAHs is not modeled, though this will be a focus for future work as this may have an impact on  PAH features in AGN hosts. There are certain proposed pathways for PAH grains to grow in the ISM that are incorporated here; during the course of a simulation, the GSD can further evolve through metal accretion, where grains grow by accumulating gas-phase interstellar matter, and coagulation, where two or more existing grains are combined. Collectively, these processes inform the dust GSD, which impacts both the luminosity and feature ratios of PAHs. Finally, the ionization fraction for PAHs also affects the relative strengths of PAH features. As in N23, we adopt the standard grain size-dependent ionization fraction from \cite{draine_2021}.

The initial conditions for our simulated galaxy are similar to the idealized Milky Way (MW) analog from N23, which uses initial conditions developed by \cite{hopkins_2011}. The galaxy is allowed to evolve until $t=9\,\mathrm{Gyr}$, at which point the distribution of PAHs is relatively constant across radii in the galaxy, and more importantly, there is a significant amount of PAHs near the nucleus ($r<0.5\,\mathrm{kpc}$). This helps when considering the impact of an AGN on PAH luminosity and PAH ratios as a function of radius in the galaxy. In this idealized setup, these qualities come at the cost of a high PAH mass fraction ($q_{\mathrm{PAH}}=M_{\mathrm{PAH}}/M_{\mathrm{dust}}$); at $t=9\,\mathrm{Gyr}$, $q_{\mathrm{PAH}}=19\%$, whereas nearby star-forming galaxies typically have $q_{\mathrm{PAH}}\approx3-6\%$ \citep[e.g.,][]{sutter_2024}. For much of our analysis, we are not concerned with $q_{\mathrm{PAH}}$ as we mostly consider the \emph{change} in PAH spectra across AGN luminosity. Where we do not deal with ratios, i.e., the impact of continuum dilution from an AGN, we also consider an earlier state of the galaxy evolution with $q_{\mathrm{PAH}}\approx6\%$ ($t=2\,\mathrm{Gyr}$). In Figure~\ref{fig:gas_density_plot}, we show the gas density distributions for both of these simulation states.

In order to most easily interpret the observed effects of the AGN on the output SED, and for simplicity, the AGN is only introduced during the postprocessing radiative transfer simulation (see \S\ref{radiative_transfer}). By only including the impact of AGN in post processing, we neglect any impact of the AGN on the galaxy's physical properties~\citep[see, e.g.,][]{Torrey2020, Sivasankaran2022, Sivasankaran2025, Sivasankaran2026}, while at the same time, we isolate the impact of the AGN's luminosity on the PAH emission. This is the goal of the present numerical experiment. 
 
\begin{figure*}
 	\centering
 	\includegraphics[width=0.95\textwidth]{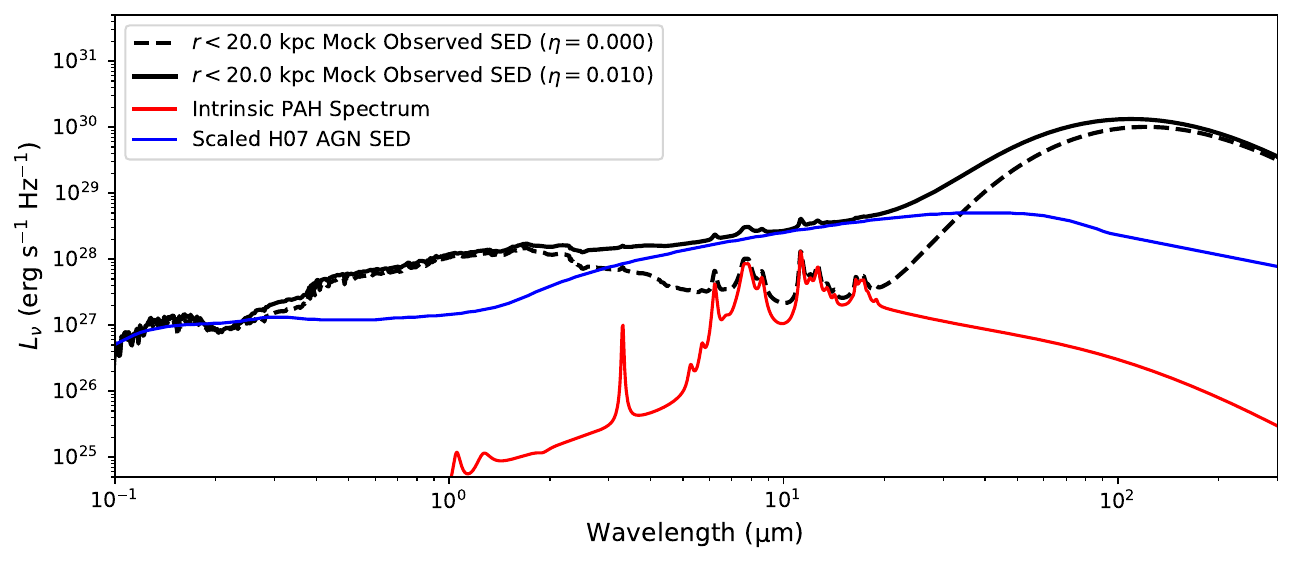}
 	\caption{
 		Example decomposed global SED for the simulated galaxy with a realistic PAH mass fraction ($q_{\mathrm{PAH}}\approx6\%$ at $t=2\,\mathrm{Gyr}$) with a moderate-luminosity AGN ($\eta=0.01$). The black line is the total SED including all luminous sources after the radiative transfer, including the AGN, and the red line shows the PAH-only spectrum after radiative transfer. The blue line is the unattenuated \cite{hopkins_2007} template AGN SED scaled to the present AGN luminosity. The SED for the same galaxy snapshot, but with no AGN ($\eta=0$), is shown as the black dashed line. Note that the PAHs have been noticeably diluted in the total SED when the AGN is present.
 	}
 	\label{fig:SED_decomp_plot}
 \end{figure*}
 
 \subsection{PAH Spectra from \texttt{POWDERDAY}} \label{radiative_transfer}

We employ the \texttt{POWDERDAY} \citep{POWDERDAY} code to facilitate the radiative transfer that results in SEDs with self-consistent PAH luminosity and feature ratios. \texttt{POWDERDAY} is built on the simulation grid generation package \texttt{YT} \citep{yt}, the radiative transfer code \texttt{HYPERION} \citep{HYPERION}, and \texttt{FSPS} \citep{FSPS} for modeling the stellar population synthesis. From the outputs of our simulated \texttt{AREPO}+\texttt{SMUGGLE} galaxy (though \texttt{POWDERDAY} is generalized to support other simulation frameworks), \texttt{POWDERDAY} compiles the ISRF for each cell in the simulation grid, where the ISRF has been reddened according to the local extinction law calculated from the cell dust GSD. A first round of Monte Carlo radiative transfer, which excludes extinction from PAHs, is then performed so that the ISRF of each cell ``sees" the photon contributions from each other cell in the simulation grid. 

\texttt{POWDERDAY} then applies the models of \cite{draine_2021}, and in particular the updates to this model from \cite{richie_2026}; these PAH models are sensitive to the intensity and hardness of the ISRF, the PAH GSD, and the distribution of ionization states, giving the PAH spectrum for each cell. The \cite{richie_2026} model leverages the single photon approximation (SPA): the assumption that PAHs always exist in the stochastic heating regime, out of equilibrium with the surrounding radiation field. This enables efficient computation of exact PAH emission spectra. When PAHs are in intense radiation field environments, multi-photon effects can impact the feature strength ratios. We discuss the SPA vs. multi-photon PAH heating in the Appendix, where we also demonstrate that the majority of cells in our model galaxy are in the single-photon regime, and we note regimes where caution must be exercised. Finally, there is a second round of radiative transfer so that extinction due to PAHs is accounted for. Thus, \texttt{POWDERDAY} self-consistently models the PAH emission within our simulated galaxy. 

We implement an AGN in our model by placing a black hole in the exact center of gas mass of the simulated galaxy, with a mass equal to that of Sgr $\mathrm{A^*}$, $M_{\mathrm{BH}}=\bhMass\,\mathrm{M_{\odot}}$ \citep{mw_bh_mass}. For our purposes, this black hole is essentially a point source with a chosen luminosity and SED shape. For the black hole SED, we use the models from \cite{hopkins_2007} (H07, see the blue line in Figure~\ref{fig:SED_decomp_plot}), which critically includes the infrared component from a hot dust torus surrounding the nucleus. The MIR spectrum of an AGN often includes broad features centered at ${\approx}10\mathrm{\mu m}$ and ${\approx}18\mathrm{\mu m}$ from silicate-rich dust grains in the torus \citep{Nenkova_I, Nenkova_II}; these features are separate from the otherwise smooth continuum and they can appear in either absorption or emission, depending on the viewing angle of the AGN geometry. For the present analysis, these would complicate the interpretation of how the AGN continuum impacts the visibility of PAH emission because they overlap with PAH features. The H07 template provides the advantage of a relatively featureless continuum, without built-in silicate features.

The AGN SED is then scaled such that its integrated power is equal to the input black hole luminosity. A useful benchmark for AGN luminosity is the Eddington luminosity, which is set by $M_{\mathrm{BH}}$,

\begin{equation} \label{eq:L_eddington}
	L_{\mathrm{Edd}} = 1.25 \times 10^{38}\, \mathrm{erg\,s^{-1}} (M_{\mathrm{BH}}/\mathrm{M_{\odot}})\,.
\end{equation}

\noindent For Sgr $\mathrm{A^*}$, $L_{\mathrm{Edd}} = 5.4\times10^{44}\,\mathrm{erg\,s^{-1}}$. In this work, we scale the AGN luminosity in terms of the Eddington ratio, $\eta$, where

\begin{equation} \label{eq:L_AGN}
	\eta = \frac{L_{\mathrm{AGN}}}{ L_{\mathrm{Edd}}}\,.
\end{equation}

\noindent Typical values for $\eta$ among nearby AGN-hosts range from ${\approx}\etaObsMin$ to ${\approx}\etaObsMax$\ \citep[e.g.,][]{gupta_2025}. We ran 10 \texttt{POWDERDAY} runs on the MW-analog galaxy between $0 \leqslant \eta \leqslant 1$, and extracted SEDs for each. An example of one of these SEDs is given in Figure~\ref{fig:SED_decomp_plot}.

This work is focused on investigating how AGN radiation affects PAH spectra, both visually and by providing photons for PAH excitation, but separately from any possible impact on the physical properties of PAHs like grain destruction or ionization. Thus, we do not include a photodestruction mechanism here. As we wish to later study the reverse, i.e., changes to PAH spectra from changes to the GSD/grain ionization absent of AGN radiative effects, a goal for future work is to incorporate a photodestruction mechanism into \texttt{POWDERDAY}. 

\section{Results} \label{sec:results}

\subsection{Interstellar Radiation Fields} \label{sec:ISRFs}

\begin{figure*}
	\centering
	\includegraphics[width=0.95\textwidth]{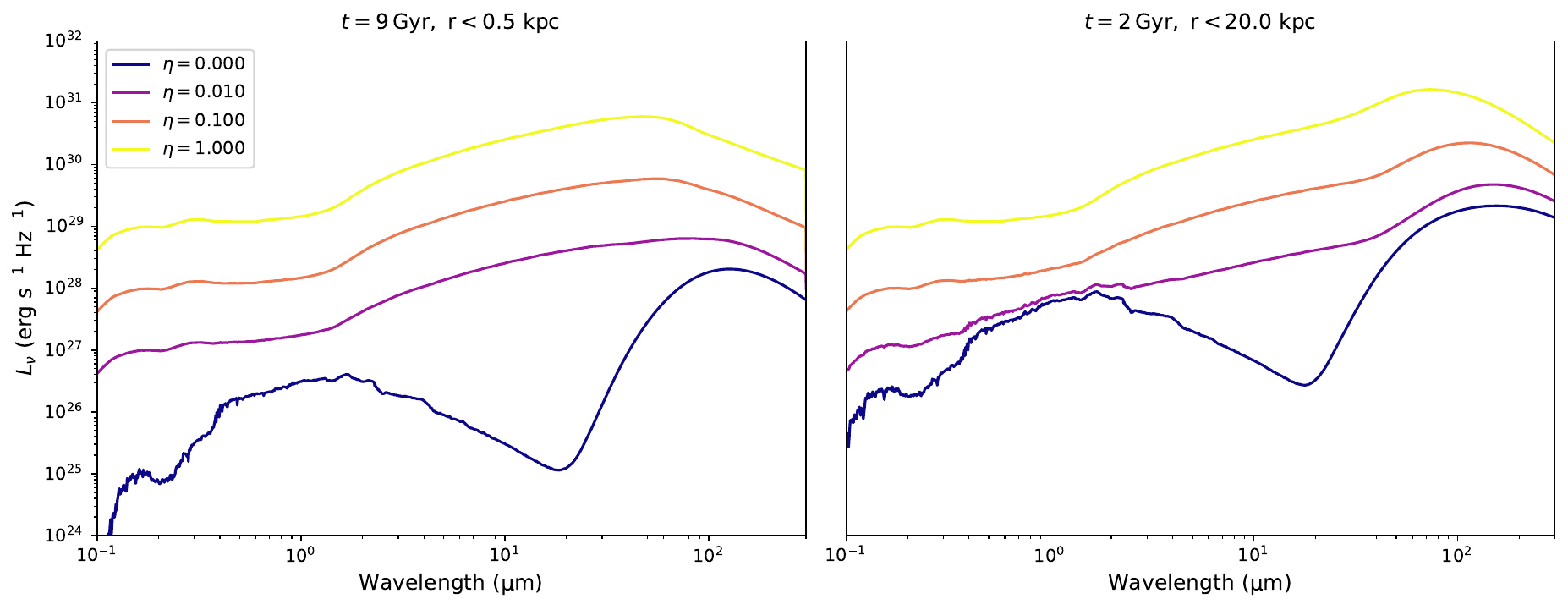}
	\caption{
	Interstellar radiation fields incident on dust grains for different values of $\eta$, integrated within the region $r<0.5\,\mathrm{kpc}$ for the galaxy at $t=9\,\mathrm{Gyr}$ (left panel), and the region $r<20\,\mathrm{kpc}$, which contains most of the simulated galaxy, for the galaxy at $t=2\,\mathrm{Gyr}$ (right panel). The AGN SED becomes the dominant part of the galaxy SED even at moderate $\eta$.}
	\label{fig:input_SEDs}
\end{figure*}

 Both the shape and the intensity of the ISRF incident upon PAHs have effects on PAH spectra under the SPA; the shape determines which photon energies PAHs are most likely to encounter, and the intensity determines by what amounts. In Figure \ref{fig:input_SEDs}, we show the ISRFs incident on dust (and therefore PAHs) at small radii $r<0.5\,\mathrm{kpc}$, and for the majority of the model galaxy $r<20\,\mathrm{kpc}$, for a few values of $\eta$. For the small radius, we use the $t=9\,\mathrm{Gyr}$ instance because there are PAHs present near the center, and we use the $t=2\,\mathrm{Gyr}$ instance for the large aperture. These ISRFs are integrated for all radii within either limit (0.5 or 20\,kpc). The smaller aperture thus simulates a small-FOV observation at the center of a nearby AGN-host, while the larger aperture is analogous to the spectrum of a distant AGN host where the host galaxy is not resolved and/or has been contaminated by the point spread function of the nuclear source. The $\eta~=~0$ curves serve as references for the case with no AGN, where there are two major components to the ISRF: the integrated stellar SED at UV/optical/NIR wavelengths, and the thermal re-emission of starlight by dust grains at IR wavelengths. It is important to note that PAHs are most selective of UV/optical photons \citep[though they are also heated by softer radiation, e.g.,][]{groves_2012, crocker_2013}, so the resulting shape of an intrinsic PAH spectrum is particularly sensitive to changes in the ISRF shape at those wavelengths.

The left panel of Figure~\ref{fig:input_SEDs} shows that when an AGN is present, the contribution to the ISRF from the AGN completely overwhelms the underlying galaxy SED even at the lowest $\eta$ values. As soon as the AGN is present, the total ISRF quickly resembles the H07 template. Further, while increases in $\eta$ scale the ISRF upward, there is little change in the shape of the ISRF except for the visible shift in the peak of the IR bump towards shorter wavelengths as the AGN increases the average temperature of the dust. In the right panel of Figure~\ref{fig:input_SEDs}, we show the ISRF contributions for the entire galaxy within $r<20\,\mathrm{kpc}$. Here, the result is similar to the nuclear aperture, though not quite as dramatic. When considering the entire galaxy, a relatively low-luminosity AGN with $\eta=0.01$ contributes a radiation field comparable in total luminosity to the host galaxy. We note that the AGN continuum already dominates the MIR spectrum even in the lowest luminosity case that we consider. As a result, we expect the AGN to significantly impact the detectability of PAH features from the mid-IR SED from galaxies hosting AGN, discussed next. 

\subsection{Observational Implications of the AGN Continuum} \label{sec:dilution}

\begin{figure*}
	\centering
	\includegraphics[width=0.99\textwidth]{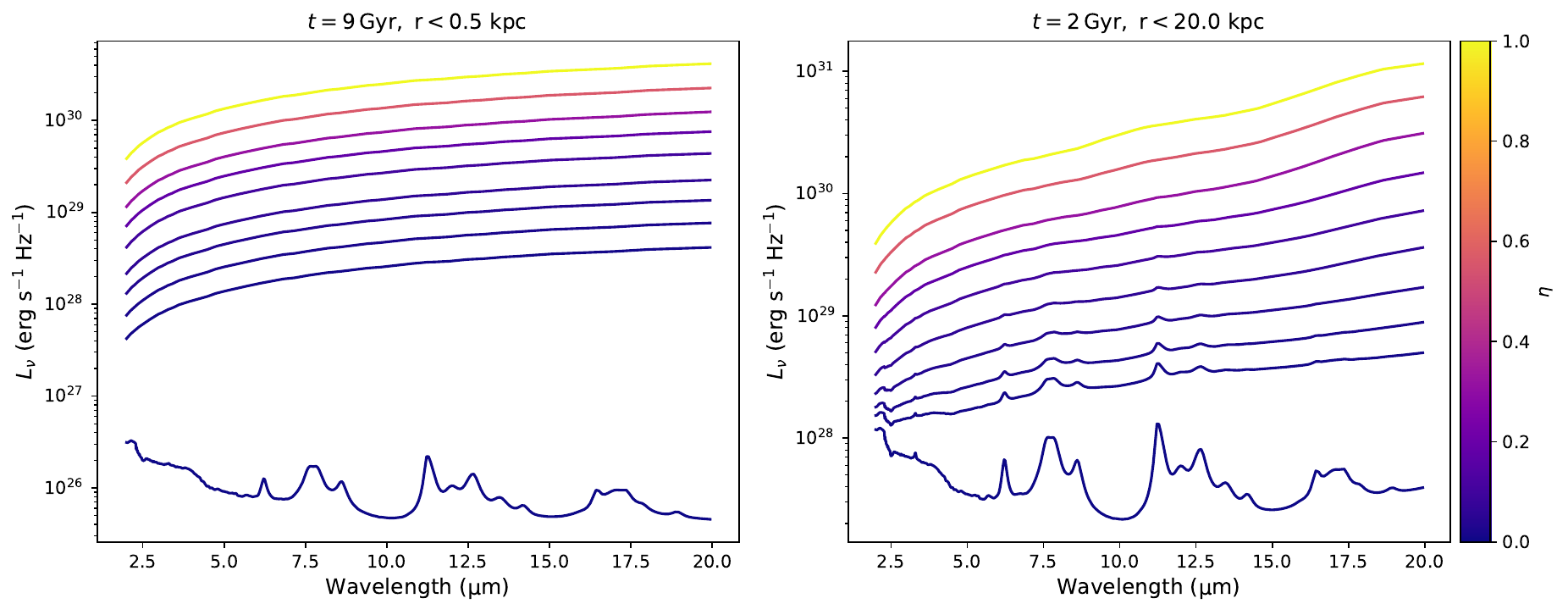}
	\caption{		
		A subset of the MIR spectra extracted from our simulated galaxy with different AGN luminosity scale factors $\eta$ at $r<0.5\,\mathrm{kpc}$ (and $t=9\,\mathrm{Gyr}$) and $r<20\,\mathrm{kpc}$ (and $t=2\,\mathrm{Gyr}$) for the left and right panels, respectively. Different colors represent the different \texttt{POWDERDAY} runs with different values of $\eta$. At low $\eta$, PAH spectra are prominent, but are reduced in prominence as the AGN luminosity increases.
	}
	\label{fig:waterfall}
\end{figure*}

\begin{figure}
	\centering
	\includegraphics[width=0.49\textwidth]{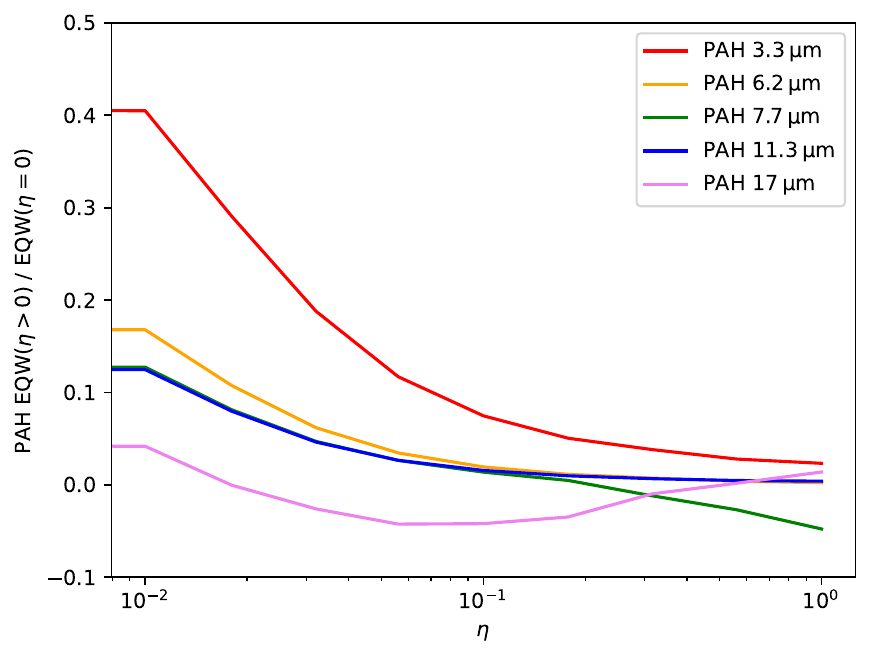}
	\caption{
        Each curve represents the fractional change in the EQW, relative to the $\eta=0$ \texttt{POWDERDAY} run (no AGN), for different PAH features as $\eta$ increases. The PAH features have been integrated over $r<20\,\mathrm{kpc}$ for the $t=9\,\mathrm{Gyr}$ simulation state, which has a more even distribution of PAHs across radii. The increased impact of dilution from the AGN continuum lowers the PAH EQW as the relative AGN luminosity increases. The $17\,\mathrm{\mu m}$ PAH feature is particularly affected by continuum dilution, and the negative EQW results from the increased curvature of the ${\sim}17\,\mathrm{\mu m}$ continuum at intermediate $\eta$.
	}
	\label{fig:EQW}
\end{figure}

We ``observe" the simulated galaxy in a face-on orientation to obtain the integrated SED within two artificial apertures centered on the AGN: one with $r<0.5\,\mathrm{kpc}$ of the $9\,\mathrm{Gyr}$ instance, and one with $r<20\,\mathrm{kpc}$ of the $2\,\mathrm{Gyr}$ instance, as in Section \ref{sec:ISRFs}. In Figure~\ref{fig:waterfall}, we show the 2---20$\mathrm{\,\mu m}$ spectra from the \texttt{POWDERDAY} radiative transfer simulations with different values of $\eta$. For both apertures, the spectrum from the run with no AGN ($\eta=0$) presents PAH features typical of a normal star-forming galaxy. However, at small radii, the MIR spectrum of even the weakest AGN in our set ($\eta=0.01$) has been completely overwhelmed by the AGN SED such that no PAHs are visible. When considering the large-aperture observations (i.e. the galaxy as a whole; the $r<20$\,kpc aperture), the continuum dilution effect from the AGN is less dramatic, and PAH emission features are visible at moderate $\eta$, though they become impossible to distinguish from the continuum for the most luminous AGN model (Figure~\ref{fig:waterfall}, right). The PAHs apparently disappear from the spectrum despite the fact that for each value of $\eta$, the galaxy has the same $q_{\mathrm{PAH}}\sim6\%$, which is typical of nearby star-forming galaxies.

The diluting impact of the AGN continuum on PAHs can be quantified using the PAH equivalent width (EQW). This can be obtained using model fit spectral decomposition tools \citep[e.g., \texttt{PAHFIT},][discussed in Section~\ref{sec:PAH_ratios}]{smith_07}, but we opt here for a simpler, model-independent method to facilitate an easier interpretation for these spectra where PAH features are difficult to quantify. For a given PAH feature, we first obtain the flux by integrating the continuum-subtracted feature profile, where the continuum is defined as a power law connecting the flanking ``clip" points described in \cite{draine_2021}. The EQW is then found by dividing the integrated flux by the point on the inferred continuum (which is in units of $f_\lambda$) that lies beneath the peak of the PAH feature. 

The fractional change in EQW as a function of $\eta$, relative to the $\eta=0$ run, is shown in  Figure~\ref{fig:EQW} for PAH features in the $r<20\,\mathrm{kpc}$ spectra (i.e., right panel of Figure~\ref{fig:waterfall}). PAH EQW decreases with increased relative AGN luminosity as the impact of continuum dilution is also increased, where the PAH EQWs are already lowered to less than half of its original value for the weakest nonzero $\eta$ run. The EQW for each feature then decreases at different rates; for the H07 AGN template shape, longer wavelength features are more affected initially, but the shorter wavelength features decrease faster with $\eta$. This is contrary to the expectation that the narrower features (e.g., 3.3, 6.2, 11.3\,$\mathrm{\mu m}$) would be less affected by a featureless continuum compared with the broader features (e.g., 7.7, 17\,$\mathrm{\mu m}$). The PAH $17\,\mathrm{\mu m}$ curve is negative at intermediate $\eta$, resulting from a combination of two factors: this PAH feature is the most affected by continuum dilution, and at these values of $\eta$ the continuum is overestimated as the far infrared bump in the galaxy SED boosts the red continuum anchor more than the blue anchor (see Figure~\ref{fig:waterfall}). 

Taken together, both panels of Figure~\ref{fig:waterfall} with Figure~\ref{fig:EQW} suggest that when observing the nuclear region of an AGN-hosting galaxy, PAH emission features are quickly overwhelmed by continuum dilution from the AGN.

\subsection{The AGN as a Source of PAH Excitation} \label{sec:PAH_excitation}

In addition to continuum dilution of PAH features, AGN can also contribute to PAH excitation in our model. Here, we consider two effects from incorporating AGN of different strengths as a source of luminosity in the radiative transfer: the impact on overall PAH luminosity, and the impact on PAH feature ratios. In Sections \ref{sec:ISRFs} and \ref{sec:dilution}, we showed how AGN emission integrates with the emission from stars and dust in the total SED of the simulated galaxy. Here, we instead analyze the ``pure PAH" spectral output, isolated from the starlight and continuum components of the SED. Therefore, the following results are unaffected by AGN continuum dilution. For both of Sections~\ref{sec:PAH_luminosity} and \ref{sec:PAH_ratios}, we analyze the PAH spectra from the $t=9\,\mathrm{Gyr}$ simulated galaxy state. As we only discuss ratios relative to the $\eta=0$ run, the specific value of $q_{\mathrm{PAH}}$ is not important here.

\subsubsection{The Impact of AGN on PAH Luminosity} \label{sec:PAH_luminosity}

\begin{figure}
	\centering
	\includegraphics[width=0.45\textwidth]{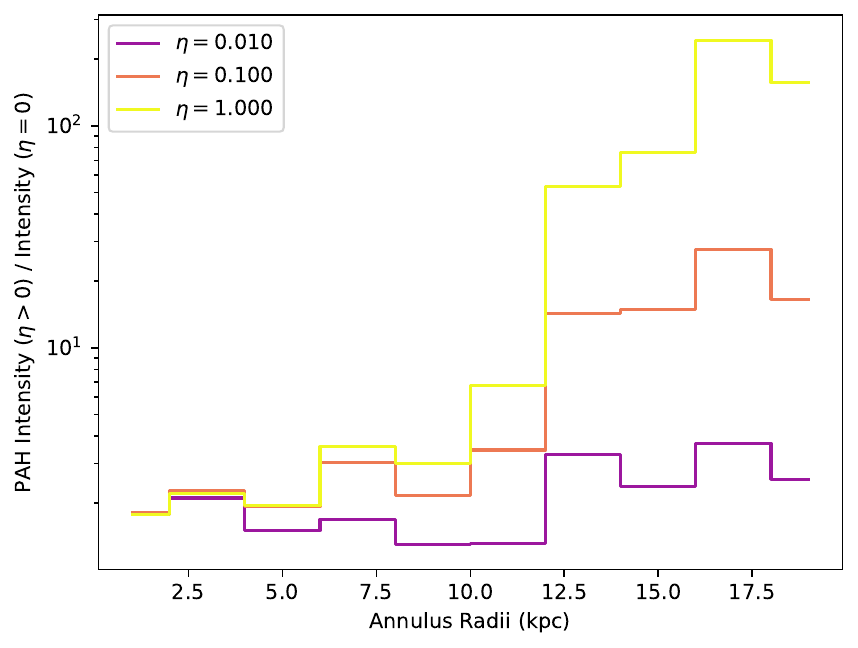}
	\caption{
	Integrated PAH luminosity within annuli for runs with $\eta>0$, normalized by the run with $\eta=0$ (no AGN), as a function of radius from the $t=9\,\mathrm{Gyr}$ galaxy center. Thus, we show the increase in PAH brightness at different radii due to photons received from the AGN. In our experiment, the AGN can contribute significantly to PAH heating even at large radii. The curves converge at small radii because of the cap of the ISRF intensity (at $\mathrm{log}\,U=2$, see Appendix~\ref{appx:SPA}) under the SPA.
	}
	\label{fig:SB_profile}
\end{figure}

Due to the contribution of exciting UV photons by the AGN (Figures~\ref{fig:SED_decomp_plot} and \ref{fig:input_SEDs}), AGN have the potential of supplying a significant fraction of the energy for PAH excitation. In Figure \ref{fig:SB_profile}, we show radial PAH surface brightness profiles for $\eta>0$ \texttt{POWDERDAY} runs, relative to the profile observed for the run with no AGN. As the only difference between the various \texttt{POWDERDAY} runs is the strength of the AGN, we can determine the radial impact of the AGN on PAH luminosity by comparing against the $\eta=0$ profile. We note that the convergence of these curves at $r\lesssim2.5\,\mathrm{kpc}$ is due to our cap placed at $\mathrm{log}\,U=2$, (see Appendix~\ref{appx:SPA}), where the innermost simulation cells may reach the peak allowed dust heating rate. 

Figure \ref{fig:SB_profile} demonstrates that the introduction of an AGN can affect PAH luminosities galaxy-wide, including at radii well separated from the nucleus; even relatively low/moderate luminosity AGN affect the surface brightness profile by boosting PAH luminosity by a factor of $\sim2$ at kiloparsec scales. For the strongest AGN that we consider ($\eta=1$), PAH surface brightness is significantly boosted at all radii in the galaxy, increasing the PAH surface brightness by two orders of magnitude at $r>12\,\mathrm{kpc}$. The PAHs at these largest radii are barely visible in the $\eta=0$ run. Thus, Figure \ref{fig:SB_profile} demonstrates the AGN is a significant or dominant source of PAH excitation even at large radii in our numerical experiment.

\subsubsection{PAH Band Ratios} \label{sec:PAH_ratios}

\begin{figure}
	\centering
	\includegraphics[width=0.45\textwidth]{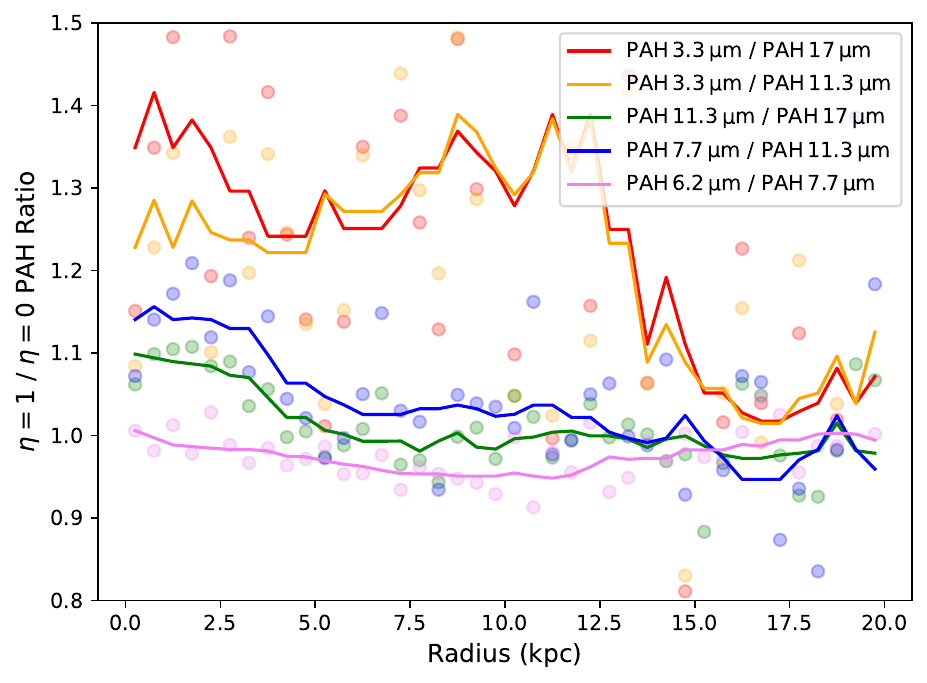}
	\caption{
	Comparison of short-to-long (STL) wavelength PAH band ratios resulting from $\eta=1$ and $\eta=0$ as a function of radius in the $t=9\,\mathrm{Gyr}$ galaxy. Data points are ratios of PAH features from PAHFIT, and lines are the data smoothed with a moving average of five data points. STL ratios are sensitive to PAH temperature, where higher STL ratios indicate hotter grains. The STL ratios tend to rise as regions approach the nucleus, showing how the AGN radiation increases the average temperature of PAH grains.
	}
	\label{fig:ratios}
\end{figure}

The observed relative strengths of PAH features can be driven by the PAH grain size distribution (GSD), ionization state, and shape and intensity of the incident ISRF. In this section, we study the impact of the AGN on the PAH feature strength ratios as driven by the AGN luminosity. For a given fixed GSD, a bluer ISRF will tend to heat PAHs to a higher temperature compared to a redder ISRF as PAHs are more likely to receive more high-energy photons. The temperature of a grain affects the relative power output between short and long wavelength PAH features, so it is useful to consider various short-to-long (STL) wavelength PAH ratios. For a fixed GSD, a bluer ISRF will produce higher STL ratios than a redder ISRF.

Similar to Section \ref{sec:PAH_luminosity}, we obtain the pure PAH spectra from a set of annuli of different radii, centered on the AGN. To quantify the emitted power within individual PAH features (so as to obtain feature ratios), rather than the total PAH luminosity, it is necessary to fit the individual features. We do this for each annulus PAH spectrum using the \texttt{PAHFIT} tool \citep{smith_07}; \texttt{PAHFIT} was designed for use with observed (e.g., Spitzer and JWST) spectra, but it can also be used to accurately fit model PAH spectra \citep[e.g.,][]{donnelly_2024, whitcomb_2024}. \citet{narayanan26a} demonstrated that \texttt{PAHFIT} accurately fits the synthetic PAH features from \texttt{POWDERDAY}. In order to best observe the effect of the AGN-powered PAHs on PAH ratios, we compare the fitted STL PAH ratios between the $\eta=0$ and the $\eta=1$ \texttt{POWDERDAY} runs. 

In Figure~\ref{fig:ratios}, we compare a range of STL PAH feature ratios as a function of galactocentric radius for the $\eta=1$ to $\eta=0$ case, using the $t=9\,\mathrm{Gyr}$ galaxy. In other words, each curve in Figure~\ref{fig:ratios} represents a short to long wavelength band ratio when comparing a model that has a luminous AGN to a model that has no AGN. Figure~\ref{fig:ratios} demonstrates that the radiation from an AGN, when used to power PAHs rather than destroy them, has the effect of boosting short wavelength features compared to long wavelength ones at nearly all locations in the model galaxy. This owes to the increased fraction of UV/optical photons contributed by the AGN SED (e.g., Figure \ref{fig:input_SEDs}). This said, not all STL ratios are affected by the same amount; STL ratios with a greater difference in wavelengths generally experience a greater boost and at larger radii. The $3.3\,\mathrm{\mu m}$ PAH feature appears particularly affected, as both ratios that involve this feature are noticeably boosted by the AGN out to $r<15\,\mathrm{kpc}$, whereas other STL ratios are only boosted up to $r<5\,\mathrm{kpc}$, if at all. Indeed, as the major PAH feature with the shortest wavelength, PAH $3.3\,\mathrm{\mu m}$ is expected to be the most sensitive to grain temperature changes. The radial profiles in Figure \ref{fig:ratios} show that an AGN can affect the PAH ratios that are often used to interpret ISM conditions in galaxies. 

\section{Discussion} \label{sec:discussion}

Using \texttt{POWDERDAY} post processing radiative transfer on a simulated MW analog, we showed that an AGN continuum can partially or totally obfuscate PAH features in a galaxy SED, and that an AGN can potentially supply photons for PAH excitation. Both of these results have implications for how PAH spectral features can be used to probe galaxy physics in galaxies where an AGN is present, and we discuss these implications in further detail in the following subsections. It is important to emphasize that this work is primarily a numerical experiment to demonstrate an instructive proof-of-concept, and that different assumptions may lead to different results. We discuss some of these assumptions here.

\subsection{Impacts of our Assumptions}

Our experiment is focused on an AGN at the center of an idealized MW analog. As demonstrated by \citet{narayanan_2023} and \citet{narayanan26a}, the physical characteristics of a galaxy can have an impact on the PAH emission characteristics in the simulated galaxy. Furthermore, the PAH characteristics and spatial distribution change within a single simulated galaxy as it evolves with time. As ISM physics dictates the evolving grain size distribution and ionization state, and the galaxy star formation history and star-dust geometry impact the incident radiation field, the total PAH power output/band ratios will be affected in turn. For instance, \cite{narayanan_2023} found that the PAH abundance was overall lower for their simulated low-metallicity dwarf galaxy analog compared with the MW analog; for this low-metallicity galaxy, the inclusion of an AGN would probably dilute PAHs at lower $\eta$. In our case, we have tried to mitigate these factors by choosing galaxy snapshots with appropriate PAH abundance for the relevant spatial apertures, and by considering ratios of PAH emission wherever possible.

Similarly, the fact that we primarily studied idealized systems may impact our results. For example, AGN are commonly found within the centers of galaxies with large bulges consisting of primarily evolved stars \citep[e.g.,][]{mcpartland_2019}. If AGN suppress star formation in the centers of galaxies through feedback \citep[for reviews, see][]{kormendy_2013, heckman_2014}, then the resulting redder central ISRFs would lead to lower local STL PAH ratios. Thus, in such a scenario, the PAH-powering effects discussed in Section \ref{sec:PAH_excitation} may be even further accentuated.

The shape of the AGN SED may also affect the impact of AGN on PAH emission features. In particular, the geometry of the region surrounding the AGN has a significant impact on the SED shape and the isotropy of emission. The composition, size, and clumpiness of the torus responsible for the bulk of AGN IR emission affects the shape of the MIR continuum and the presence and strength of silicate features. The AGN orientation may also be a factor; the viewing angle of the torus from Earth determines whether silicates are seen in emission or absorption \citep{pier_1992, pier_1993, siebenmorgen_2005}, differently affecting PAH visibility. Similarly, the viewing angle of the circumnuclear environment from different places within the AGN galaxy affects the ratio of hard to soft photons from the AGN that PAHs in a given region are exposed to. We selected the H07 template for this work as it has three important characteristics: it has significant power at UV/optical wavelengths, it has a significant MIR continuum, and it does not have silicate emission/absorption features which would make drawing broad conclusions difficult for this initial experiment. However, future work should investigate different AGN SED shapes.

\subsection{Continuum Dilution of PAHs}

In Figures~\ref{fig:waterfall} and \ref{fig:EQW}, we show how an AGN with different luminosities can alter the observed PAH spectrum of a MW-like galaxy, even though the PAH abundance and grain properties remain unchanged. Due to the effect of continuum dilution, PAH features become difficult to distinguish at high AGN luminosity even for large radius apertures, and PAH features are essentially impossible to distinguish from the continuum at small radii in our experiment. Thus, derived PAH abundances from spectra are in danger of being systematically underestimated in unresolved AGN-hosts and in the central regions of nearby AGN-hosts.  We also noted that the central wavelength and the spectral shape of a PAH feature are related to its detectability at large $\eta$. Together with the fact that PAH features are differently affected by the silicate emission bumps, this implies that the AGN continuum also stands to impact the way that PAH ratios are quantified, in addition to the total PAH emission. 

The apparently diminished total PAH luminosity and unusual PAH band ratios that are frequently observed in AGN-hosts are the most important pieces of evidence supporting the idea that the nuclear radiation destroys PAHs, and that it probably does so preferentially to smaller grains \citep[e.g.,][]{smith_07, odowd_09, sales_10, garciabernete_22, lai_2022}. Given the frequency of AGN in galaxies, especially at higher redshifts, this proposed mechanism is critical to understand in order to gain a complete picture of how the PAH GSD evolves in the ISM. Our results underscore the importance of carefully removing the AGN contribution to MIR spectra when quantifying PAH emission. This may take the form of a fit-able AGN component when using spectral decomposition tools such as \texttt{PAHFIT} \citep{smith_07} or \texttt{CAFE} \citep{marshall_2007, diaz-santos_2025}. However, this component depends on the AGN/dust geometry of a system that is typically poorly constrained, and is further complicated by the fact that the AGN impact is not entirely separable from the total SED. Thus, our simulation framework complements observational efforts by providing a means to impose continuum dilution on simulated MIR spectra for a better basis of comparison. In future work, more detailed simulations could be used to inform procedures for removing the AGN component from observed MIR spectra. 

\subsection{AGN-powered PAHs}

Our investigation of PAH luminosity at different radii showed that an AGN can be a significant source of power for PAHs, even far from the nucleus at kiloparsec scales. In our experiment, AGN radiation impinges on PAHs at these radii, implying that PAHs surviving at kiloparsec scales in real galaxies also see and therefore can be powered by AGN emission. Calibrations using PAH luminosity as a tracer for the SFR, molecular gas content or gas-phase metallicity could be skewed for galaxies where an AGN is an additional power source for PAHs. Additionally, emission from PAHs surviving closer in to the nucleus could be mistaken as a signpost for nuclear star-formation. 

Galaxies hosting an AGN are typically observed to have lower STL PAH ratios, likely indicating alterations to the PAH GSD or ionization distribution made by the AGN. However, PAHs, including smaller grains, can survive in the vicinity of an AGN due to shielding by the circumnuclear torus and surrounding molecular material. We have shown how AGN radiation reaching these surviving grains (either directly or indirectly) might alter PAH ratios in the opposite way by increasing STL ratios. 

Using spatially-resolved MIR spectral maps from Spitzer/IRS, \cite{donnelly_2024} showed radial profiles for various STL PAH ratios in the low-luminosity AGN NGC\,4138. All four of the STL ratios they investigated were suppressed within the inner kiloparsec compared with normal galaxies. However, for three of the STL ratios, the innermost $r\lesssim270\,\mathrm{pc}$ presented slightly increased values compared with the minimum suppressed ratios, potentially indicating that the AGN was boosting those ratios by partially powering PAHs in the nucleus. In Figure \ref{fig:ratios}, we see similar behavior in our simulated galaxy; larger STL ratios as radius decreases. We note, however, that PAH ratios at very small radii ($\lesssim0.5\,\mathrm{kpc}$) are impacted by the SPA. As very high radiation field intensities are expected here due to the AGN (see Figure \ref{fig:logU_dist}), multi-photon heating effects are non-negligible. This would likely have the effect of boosting power in the long-wavelength features, especially $\mathrm{PAH\ 17\,\mu m}$. Indeed, this PAH feature appears to be stronger in AGN galaxies compared with normal galaxies \citep{smith_07}.

In our model, the AGN is a luminous point source that contributes its SED to the PAH-heating ISRF, in much the same way that an individual star would in a lower amount. The resulting effect is an excess of UV/optical radiation that can increase some of the STL PAH ratios, especially at small radii. These increased STL PAH ratios from AGN-powered PAHs would somewhat offset the effect of AGN-driven small grain destruction, which lowers STL ratios. Thus, it is important to understand and quantify this radiative effect in order to properly interpret the observational signatures of grain photodestruction. Here, we have studied the effect of PAH heating by an AGN in isolation, and future work should combine this with grain processing mechanisms. Additionally, more work should be done with observed galaxies to holistically quantify the ISRF incident on PAHs in AGN galaxies, so that these observational signatures of grain processing can be understood unambiguously.

\section{Summary} \label{sec:summary}

It is difficult to interpret the physics affecting observed PAH features in AGN galaxies because multiple competing factors are at play. Among these are AGN-driven grain destruction or ionization, possible PAH excitation by an AGN, and observational effects like continuum dilution. However, by using numerical experiments/simulations such as the one outlined here, we can investigate each of these factors separately. In this work, we use the \texttt{POWDERDAY} radiative transfer code to investigate two of these aspects related to PAH spectra in a simulated galaxy with an AGN: the visual effect of AGN continuum dilution, and effects from the AGN as a source of power for PAHs. We find that the continuum originating from the AGN is capable of obfuscating PAH emission and altering apparent PAH feature ratios, especially in the central regions of the galaxy. Additionally, we find that an AGN is capable of supplying photons for PAH emission, even at large radii from the nucleus. As the AGN SED dominates the ISRF for strong AGN, this can have the effect of altering PAH ratios by shifting power to shorter wavelength features. In future work, we can apply these results when combined with processes that affect the intrinsic properties of PAH grains, such as photodestruction by AGN radiation.

\section*{Acknowledgments} 

This work originated from a project of the Summer Program in Astrophysics 2025 held at the University of Virginia, and funded by the Center for Global Inquiry and Innovation, the National Science Foundation (Grant 2452494), the National Radio Astronomy Observatory (NRAO), the Kavli Foundation and the Heising-Simons Foundation. D.N. was funded by NASA via grants ATP21-0013, ATP23-0002, and ADSPS23-0007. The authors acknowledge Research Computing at The University of Virginia for providing computational resources and technical support that have contributed to the results reported within this publication  (URL: \href{https://rc.virginia.edu}{https://rc.virginia.edu}). PT acknowledges support from the National Science Foundation under Cooperative Agreement 2421782 and the Simons Foundation grant MPS-AI-00010515 awarded to NSF-Simons AI Institute for Cosmic Origins -- CosmicAI, \href{https://www.cosmicai.org/}{https://www.cosmicai.org/}. HL is supported by the National Key R\&D Program of China No. 2023YFB3002502, the National Natural Science Foundation of China under No. 12373006 and 12533004, and the China Manned Space Program with grant No. CMS-CSST-2025-A10.

\appendix 
\restartappendixnumbering

\section{Single vs. Multi-photon PAH Heating} \label{appx:SPA}

In most interstellar environments, the small absorption cross sections of PAHs allow them to be stochastically heated; that is, PAHs absorb photons one-at-a-time and all absorbed energy from one absorption event has been re-radiated before the next photon is absorbed. However, when PAHs are in or near extreme radiation environments such as star-forming regions, this may not be true. In this context, a useful metric for the intensity of the radiation field is the dust heating parameter $U$; as described in \cite{draine_2021},

\begin{equation} \label{eq:U}
	U \equiv \frac{1}{h_{\mathrm{ref}}} \int d\nu\ u_{*,\nu}\ c\ C_{\mathrm{abs}}^{\mathrm{Ad}}(\nu) \,,
\end{equation}

\noindent where $\nu$ is the photon frequency, $c$ is the speed of light, and $C_{\mathrm{abs}}^{\mathrm{Ad}}(\nu)$ is the frequency-dependent absorption cross-section of a grain. $U$ is normalized with respect to the ISRF in the vicinity of the Sun \citep[i.e., a modified version of the ISRF of][]{MMP}, thus $u_{*,\nu}$ and $h_{\mathrm{ref}}$ are a reference energy density and a reference dust heating rate for the \cite{MMP} ISRF, respectively. By definition, $\mathrm{log}\,U=0$ for the intensity and shape of the ISRF in the vicinity of the Sun. 

Single photon heating generally occurs at $\mathrm{log}\,U<2.5$, above which multi-photon heating is expected to become significant \citep{draine_2021}. An important detail lies in the fact that the threshold value of $\mathrm{log}\,U$ for a given PAH depends on the PAH size, as larger PAHs have larger absorption cross-sections. Indeed, the largest PAHs experience multi-photon heating at $\mathrm{log}\,U\sim2$.

PAH feature ratios are determined in part by the grain temperatures; hotter PAHs emit more at short wavelengths and vice versa. The practical upshot of single vs. multi-photon heating is that for the former, a fixed-shape ISRF does not impact PAH ratios as the ISRF \textit{intensity} increases, as long as $\mathrm{log}\,U<2.5$, because the isolated photon absorption events are still drawing from the same flux-weighted statistical distribution of incident photon energies. However, while in the multi-photon absorption regime, each additional absorbed photon increases the temperature of a grain, thus PAH band ratios depend on $U$, even for a fixed ISRF shape. Longer-wavelength PAH bands tends to increase first, because these bands are primarily carried by larger PAHs with larger cross-sections. 

Recently, \cite{richie_2026} showed that an accurate PAH spectrum could be quickly computed for any \textit{arbitrary} shape of the incident ISRF where PAHs are in the single-photon heating regime. This single photon assumption (SPA) has enabled the calculation of more accurate PAH spectrum shapes from \texttt{POWDERDAY} where the SPA is valid. In this work, we run \texttt{POWDERDAY} using the SPA formulation from \cite{richie_2026} for this reason. Because we apply the SPA, there is a cap on the dust heating rate within a given simulation cell at $\mathrm{log}\,U=2$.

\begin{figure}
	\centering
	\includegraphics[width=0.45\textwidth]{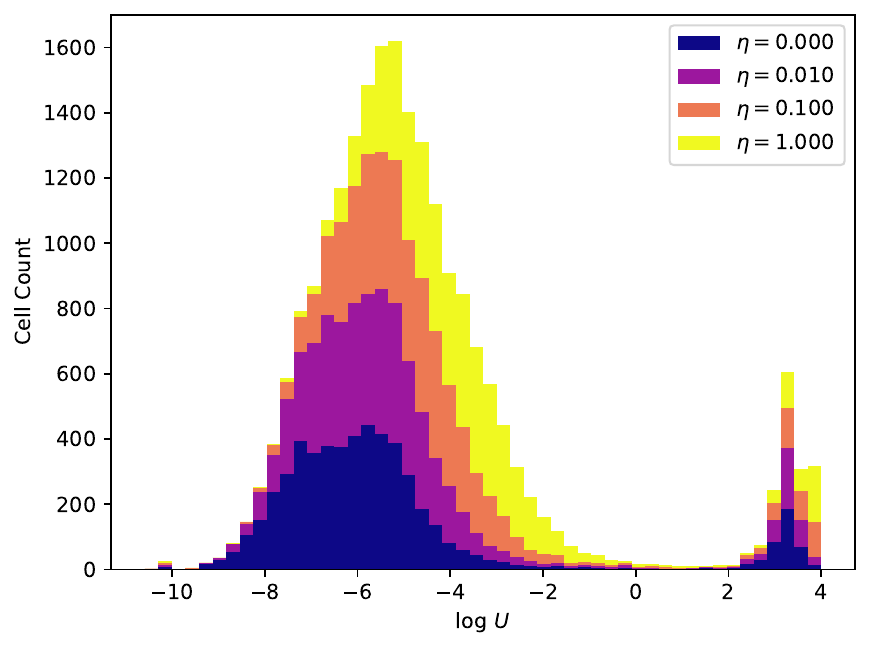}
	\caption{
		Stacked histograms showing the distributions of the dust heating parameter $U$ for \texttt{POWDERDAY} runs of various AGN intensities (scaled by $\eta$), for simulation cells with $r<20\,\mathrm{kpc}$. The adopted assumption that PAHs are heated stochastically, valid for $\mathrm{log}\,U\lesssim2.5$, is applicable for the majority of the galaxy.
	}
	\label{fig:logU_dist}
\end{figure}

To ensure that the SPA generally holds, we inspect the computed $\mathrm{log}\,U$ distributions from various \texttt{POWDERDAY} runs with multi-photon heating enabled across different $\eta$; these are shown in Figure \ref{fig:logU_dist}. For all tested AGN intensities, the majority of cells in the simulated galaxy have $\mathrm{log}\,U\ll3$, indicating that the SPA is generally appropriate. Indeed, most cells exhibit very low $U$, much less than the archetypical $\mathrm{log}\,U=0$, likely owing to the relatively low resolution of our particular \texttt{AREPO} simulation. The whole $U$ distribution moves towards larger values for larger values of $\eta$, indicating that the AGN is contributing to dust heating throughout the galaxy. The smaller peak centered at $\mathrm{log}\,U\sim3$ is likely driven primarily by star-forming regions, given that it is visible also when $\eta=0$. However, this smaller peak has a wing at higher $\mathrm{log}\,U$ for $\eta=1$, suggesting very high dust temperatures in the cells closest to the AGN. Thus, we must exercise caution in interpreting PAH band ratios at these small radii. 

\newpage
\bibliography{citations}{}
\bibliographystyle{aasjournalv7}

\end{CJK*}
\end{document}

%% file: values.tex
\newcommand\bhMass{4.3\times10^{6}} 
\newcommand\etaObsMin{10^{-3}} 
\newcommand\etaObsMax{1} 